\documentclass[epj]{webofc}
\usepackage[utf8]{inputenc}
\usepackage[varg]{txfonts}   
\usepackage{booktabs}
\usepackage{xcolor}
\definecolor{darkred}{rgb}{0.4,0.0,0.0}
\definecolor{darkgreen}{rgb}{0.0,0.4,0.0}
\definecolor{darkblue}{rgb}{0.0,0.0,0.4}
\usepackage[bookmarks,linktocpage,colorlinks,
    linkcolor = darkred,
    urlcolor  = darkblue,
    citecolor = darkgreen]{hyperref}
\usepackage{subfigure}
\usepackage{bm}
\usepackage{amssymb}
\wocname{EPJ Web of Conferences}
\woctitle{Lattice2017}
\begin{document}
%
\selectlanguage{english}
\title{
Baryon interactions from lattice QCD with physical masses ----- strangeness 
$\bm{S=-1}$ sector -----
}
\author{%
\firstname{Hidekatsu} \lastname{Nemura}\inst{1,2}\fnsep\thanks{Speaker, \email{hidekatsu.nemura@rcnp.osaka-u.ac.jp}} \and
\firstname{Sinya} \lastname{Aoki}\inst{2,3,4} \and
\firstname{Takumi} \lastname{Doi}\inst{2,5} \and
\firstname{Shinya} \lastname{Gongyo}\inst{2} \and
\firstname{Tetsuo} \lastname{Hatsuda}\inst{2,5} \and
\firstname{Yoichi} \lastname{Ikeda}\inst{1,2} \and
\firstname{Takashi} \lastname{Inoue}\inst{2,6} \and
\firstname{Takumi} \lastname{Iritani}\inst{2} \and
\firstname{Noriyoshi} \lastname{Ishii}\inst{1,2} \and
\firstname{Takaya} \lastname{Miyamoto}\inst{2,3} \and
\firstname{Kenji} \lastname{Sasaki}\inst{2,3} 
%
}
\institute{%
 Research Centre for Nuclear Physics, Osaka University, Osaka, 567-0047, Japan
\and
 Theoretical Research Division, Nishina Centre, RIKEN, 
 Saitama, 351-0198, Japan
\and
 Centre for Gravitational Physics, 
 Yukawa Institute for Theoretical Physics, Kyoto University, 
 Kyoto, 606-8502, Japan
\and
 Centre for Computational Sciences, University of Tsukuba, Tsukuba 305-8577, Japan
\and
 iTHEMS Program and iTHES Research Group, RIKEN, Saitama, 351-0198, Japan
\and
 Nihon University, College of Bioresource Sciences, 
 Kanagawa 252-0880, Japan
%
}
\abstract{%
 We present our recent results of baryon interactions with strangeness 
 $S=-1$ based on Nambu-Bethe-Salpeter (NBS) correlation functions 
 calculated from lattice QCD with almost physical quark masses 
 corresponding to 
 $(m_\pi,m_K)\approx(146,525)$ MeV and large volume 
 $(La)^4=(96a)^4\approx$ (8.1 fm)$^4$. 
 In order to perform a comprehensive study of baryon interactions, 
 a large number of NBS correlation functions from NN to $\Xi\Xi$ are 
 calculated simultaneously by using large scale computer resources. 
 In this contribution, 
 we focus on the strangeness $S=-1$ channels of the hyperon
 interactions by means of HAL QCD method. 
 Four sets of three potentials 
 (the $^3S_1-^3D_1$ central, $^3S_1-^3D_1$ tensor, and 
 the $^1S_0$ central potentials) are presented for 
 the $\Sigma N - \Sigma N$ (the isospin $I=3/2$) diagonal, 
 the $\Lambda N - \Lambda N$ diagonal, 
 the $\Lambda N \rightarrow \Sigma N$ transition, and 
 the $\Sigma N - \Sigma N$ ($I=1/2$) diagonal interactions. 
 Scattering phase shifts for $\Sigma N$ $(I=3/2)$ system are presented. 
%
 \\{\normalsize\vspace*{-44.0em}\begin{flushright}
 YITP-17-121,RIKEN-QHP-341,RIKEN-ITHEMS-REPORT-17
 \end{flushright}\vspace*{39.72em}}
}
\maketitle
\section{Introduction}

Nuclear force and strangeness nuclear forces provide an important starting 
point to understand how hypernuclei are bound, in which hyperons 
(or strange quarks) are embedded in normal nuclei as 
``impurities''\cite{Hashimoto:2006aw}. 
Determining how such a baryon-baryon interaction is 
described from a fundamental 
perspective is a challenging problem in physics. 
Although a normal nucleus is successfully described by utilising the high 
precision nucleon-nucleon ($NN$) potentials together with a three-nucleon 
force a quantitatively same-level description of a hypernucleus is still 
difficult because of large uncertainties of hyperon-nucleon ($YN$) and 
hyperon-hyperon ($YY$) interactions; 
those $YN$ and $YY$ potentials are not well constrained from experimental 
data due to the short life time of hyperons. 
A recent experimental study shows a tendency to repulsive $\Sigma$-nucleus 
interaction
and 
only a four-body $\Sigma$-hypernucleus ($^4_{\Sigma}$He) has 
been observed; 
those suggests a repulsive nature of the $\Sigma N$ interaction. 
It has been pointed out that a $\Lambda N-\Sigma N$ coupled-channel 
interaction accompanied with $^3S_1-^3D_1$ mixing by tensor operator 
plays a vital role to have a hypernucleus being bound\cite{Nemura:2002fu}. 
Such quantitative understanding is useful to study properties of 
hyperonic matters inside the neutron stars, 
where recent observations of massive neutron star heavier than 
$2M_{\odot}$~\cite{Demorest:2010bx,Antoniadis:2013pzd} may be issued 
against a hyperonic equation of state (EOS) employed in such a study. 
Furthermore, 
better 
understanding of $YN$ and $YY$ is becoming increasingly important 
due to the observation of the 
binary neutron star merger\cite{TheLIGOScientific:2017qsa,GBM:2017lvd}.

During the last decade 
a new lattice QCD approach to study a hadron-hadron interaction 
has been proposed\cite{Ishii:2006ec,Aoki:2009ji} and 
developed to overcome the numerical difficulty\cite{HALQCD:2012aa}. 
In this approach, the interhadron potential is obtained 
by means of the lattice QCD measurement of the Nambu-Bethe-Salpeter (NBS) 
wave function. 
The observables such as the phase shifts and the binding energies are 
calculated by using the resultant potential\cite{Aoki:2012tk}. 
A large scale lattice QCD calculation is 
now in progress\cite{DoiIshiiSasaki2017LAT} to study the baryon 
interactions from $NN$ to $\Xi\Xi$ 
by measuring 
the NBS wave functions for 
52 channels from the $2+1$ flavor lattice QCD. 
See also Ref.\cite{Gongyo:2017fjb} for the study of $\Omega\Omega$ interaction.

The purpose of this report is to present 
our recent results of the 
$\Lambda N-\Sigma N$ (both the isospin $I=1/2,3/2$) systems 
using full QCD gauge configurations. 
Several earlier results had already been reported 
at LATTICE 2008, 
LATTICE 2009 
 and 
LATTICE 2011\cite{Nemura:2012fm} 
with heavier quark masses and smaller lattice volumes. 
Although the possibility of ``mirage'' is pointed out\cite{Iritani:2016jie}, 
calculations with larger quark masses 
for the $\Sigma^{-}n$ channel are 
found in Ref.\cite{Beane:2012ey}. 
This report shows the latest results of those studies, 
based on recent works reported at 
LATTICE 2013\cite{Nemura:2014eta,Nemura:2015yha};
the baryon-baryon interaction in the strangeness $S=-1$ sector 
(i.e, $\Lambda N-\Lambda N$, $\Lambda N-\Sigma N$, and 
$\Sigma N-\Sigma N$ (both $I=1/2$ and $3/2$)) is studied 
at almost physical quark masses 
corresponding to ($m_{\pi}$,$m_{K}$)$\approx$(146,525)MeV and 
large volume $(La)^4=(96a)^4\approx$ (8.1 fm)$^4$. 

\section{Outline of the HAL QCD method}

In order to study the baryon-baryon interactions, 
we first define the equal time NBS wave function 
in particle channel $\lambda=\{B_{1},B_{2}\}$ 
with Euclidean time $t$~\cite{Ishii:2006ec,Aoki:2009ji} 
\begin{equation}
 \phi_{\lambda E}(\vec{r}) {\rm e}^{-E t} = 
 \sum_{\vec{X}}
 \left\langle 0
  \left|
   B_{1,\alpha}(\vec{X}+\vec{r},t)
   B_{2,\beta}(\vec{X},t)
  \right| B=2, E, S, I 
 \right\rangle,
  \label{DefineNBSWF}
\end{equation}
where 
$B_{1,\alpha}(x)$ ($B_{2,\beta}(x)$) denotes the local interpolating field of 
baryon $B_{1}$ ($B_{2}$) 
with mass $m_{B_{1}}$ ($m_{B_{2}}$), 
and 
$E=\sqrt{k_{\lambda}^2+m_{B_{1}}^2}+\sqrt{k_{\lambda}^2+m_{B_{2}}^2}$ 
is the total energy 
in the centre of mass system of a baryon number $B=2$, strangeness $S$, 
and isospin $I$ state. 
For $B_{1,\alpha}(x)$ and $B_{2,\beta}(x)$, 
we employ the local interpolating field of octet baryons given by 
%
\begin{equation}
 \!\!\!
 \begin{array}{llll}
  p \! = \! \varepsilon_{abc} \left(
			 u_a C\gamma_5 d_b
			\right) u_c,\!
  &
  n \! = \! - \varepsilon_{abc} \left(
			   u_a C\gamma_5 d_b
			  \right) d_c,\!
  &
  \Sigma^{+} \! = \! - \varepsilon_{abc} \left(
				    u_a C\gamma_5 s_b
				   \right) u_c,\!
  &
  \Sigma^{-} \! = \! - \varepsilon_{abc} \left(
				    d_a C\gamma_5 s_b
				   \right) d_c,\!
  \\
  \Sigma^{0} \! = \! {1\over\sqrt{2}} \left( X_u \! - \! X_d \right),\!
  &
  \Lambda \! = \! {1\over \sqrt{6}} \left( X_u \! + \! X_d \! - \! 2 X_s \right),\!
  &
  \Xi^{0} \! = \! \varepsilon_{abc} \left(
                               u_a C\gamma_5 s_b
                              \right) s_{c},\!

  &
  \Xi^{-} \! = \! - \varepsilon_{abc} \left(
                                 d_a C\gamma_5 s_b
                                \right) s_{c},\!
  \\
  \mbox{where}
  &
  X_u = \varepsilon_{abc} \left( d_a C\gamma_5 s_b \right) u_c, 
  &
  X_d = \varepsilon_{abc} \left( s_a C\gamma_5 u_b \right) d_c,
  &
  X_s = \varepsilon_{abc} \left( u_a C\gamma_5 d_b \right) s_c.
 \end{array}
 \label{BaryonOperatorsOctet}
\end{equation}
%
For simplicity, we have suppressed the explicit spinor indices and 
spatial coordinates in Eq.~(\ref{BaryonOperatorsOctet}) and the 
renormalisation factors in Eq.~(\ref{DefineNBSWF}). 
Based on a set of the NBS wave functions, we define a non-local 
potential 
%
$\left(
   \frac{\nabla^2}{2\mu_{\lambda}} + \frac{k_{\lambda}^{2}}{2\mu_{\lambda}}
  \right)
  \delta_{\lambda \lambda^{\prime}}
  \phi_{\lambda^{\prime} E}(\vec{r}) = 
  \int d^3r^\prime\, U_{\lambda\lambda^{\prime}}(\vec{r},\vec{r^{\prime}}) 
  \phi_{\lambda^{\prime} E}(\vec{r^{\prime}})$ 
with the reduced mass 
$\mu_{\lambda}=m_{B_{1}}m_{B_{2}}/(m_{B_{1}}+m_{B_{2}})$. 

In lattice QCD calculations, 
we compute the 
four-point correlation function defined by\cite{HALQCD:2012aa} 
\begin{eqnarray}
 {F}_{\alpha\beta,JM}^{\langle B_1B_2\overline{B_3B_4}\rangle}(\vec{r},t-t_0) 
 && = 
 \sum_{\vec{X}}
 \left\langle  0 
  \left|
   B_{1,\alpha}(\vec{X}+\vec{r},t)
   B_{2,\beta}(\vec{X},t)
   \overline{{\cal J}_{B_{3} B_{4}}^{(J,M)}(t_0)}
  \right|  0 
 \right\rangle,
\end{eqnarray}
where 
$\overline{{\cal J}_{B_3B_4}^{(J,M)}(t_0)}=
  \sum_{\alpha^\prime\beta^\prime}
  P_{\alpha^\prime\beta^\prime}^{(J,M)}
  \overline{B_{3,\alpha^\prime}(t_0)}
  \overline{B_{4,\beta^\prime}(t_0)}$
is a source operator that creates $B_3B_4$ 
states with the
total angular momentum $J,M$. 
The normalised four-point function 
can be expressed as
\begin{eqnarray}
 &&
  {R}_{\alpha\beta,JM}^{\langle B_1B_2\overline{B_3B_4}\rangle}(\vec{r},t-t_0) 
  =
  {\rm e}^{(m_{B_1}+m_{B_2})(t-t_0)} 
  {F}_{\alpha\beta,JM}^{\langle B_1B_2\overline{B_3B_4}\rangle}(\vec{r},t-t_0) 
  \nonumber
  \\
  \!\!\!\!&=&\!\!\!\!
   \sum_{n} A_{n}
   \sum_{\vec{X}}
   \left\langle 0
    \left|
     B_{1,\alpha}(\vec{X}+\vec{r},0)
     B_{2,\beta}(\vec{X},0)
    \right| E_{n} 
   \right\rangle
   {\rm e}^{-(E_{n}-m_{B_1}-m_{B_2})(t-t_0)}
   \!+\! O({\rm e}^{-(E_{\rm th}-m_{B_{1}}-m_{B_{2}})(t-t_{0})}),
\end{eqnarray}
where $E_n$ ($|E_n\rangle$) is the eigen-energy (eigen-state)
of the six-quark system 
and 
$A_n = \sum_{\alpha^\prime\beta^\prime} P_{\alpha^\prime\beta^\prime}^{(JM)}$
$\langle E_n | \overline{B}_{4,\beta^\prime}
\overline{B}_{3,\alpha^\prime} | 0 \rangle$. 
Hereafter, the spin and angular momentum subscripts are suppressed 
for $F$ and $R$ for simplicity. 
At moderately large $t-t_0$ 
where the 
inelastic contribution 
above the pion production 
$O({\rm e}^{-(E_{\rm th}-m_{B_{1}}-m_{B_{2}})(t-t_{0})})=
O({\rm e}^{-m_{\pi}(t-t_{0})})$ 
becomes 
negligible, 
we can construct the non-local potential $U$ through 
$\left(
   \frac{\nabla^2}{2\mu_{\lambda}} + \frac{k_{\lambda}^{2}}{2\mu_{\lambda}}
  \right)
  \delta_{\lambda \lambda^{\prime}}
  F_{\lambda^{\prime}}(\vec{r}) = 
  \int d^3r^\prime\, U_{\lambda \lambda^{\prime}}(\vec{r},\vec{r^{\prime}}) 
  F_{\lambda^{\prime}}(\vec{r^{\prime}}).$ 
In lattice QCD calculations in a finite box, it is practical to use 
the velocity (derivative) expansion, 
$U_{\lambda \lambda^{\prime}}(\vec{r},\vec{r^{\prime}}) =
 V_{\lambda \lambda^{\prime}}(\vec{r},\vec{\nabla}_{r})
\delta^{3}(\vec{r} - \vec{r^{\prime}}).$ 
In the lowest few orders we have 
\begin{equation}
V(\vec{r},\vec{\nabla}_{r}) = 
V^{(0)}(r) + V^{(\sigma)}(r)\vec{\sigma}_{1} \cdot \vec{\sigma}_{2} + 
V^{(T)}(r) S_{12}
 + 
V^{(^{\ LS}_{ALS})}(r) \vec{L}\cdot (\vec{\sigma}_{1}\pm\vec{\sigma}_{2})
 + 
O(\nabla^{2}),
\end{equation}
where $r=|\vec{r}|$, $\vec{\sigma}_{i}$ are the Pauli matrices acting 
on the spin space of the $i$-th baryon, 
$S_{12}=3
(\vec{r}\cdot\vec{\sigma}_{1})
(\vec{r}\cdot\vec{\sigma}_{2})/r^{2}-
\vec{\sigma}_{1}\cdot
\vec{\sigma}_{2}$ is the tensor operator, and 
$\vec{L}=\vec{r}\times (-i \vec{\nabla})$ is the angular momentum operator. 
The first three-terms constitute the leading order (LO) potential while 
the fourth term corresponds to the next-to-leading order (NLO) potential. 
By taking the non-relativistic approximation, 
$E_{n} - m_{B_{1}} - m_{B_{2}} \simeq 
 {k_{\lambda,n}^{2} \over {2\mu_{\lambda}}} +
 O(k_{\lambda,n}^{4})$, 
and neglecting the $V_{\rm NLO}$ and the higher order terms, 
we obtain 
$\left(\frac{\nabla^2}{2\mu_{\lambda}} -\frac{\partial}{\partial t}\right)
{ R}_{\lambda\varepsilon}(\vec r,t)
\simeq 
V^{\rm (LO)}_{\lambda \lambda^{\prime}}(\vec{r}) 
\theta_{\lambda \lambda^{\prime}}
{ R}_{\lambda^{\prime}\varepsilon}(\vec r,t)$, with 
$\theta_{\lambda \lambda^{\prime}}=
{\rm e}^{( m_{B_{1}}+m_{B_{2}}-m_{B_{1}^{\prime}}-m_{B_{2}^{\prime}})(t-t_0)}$.
Note that we have introduced the matrix form 
${ R}_{\lambda^{\prime}\varepsilon} =
 \{R_{\lambda^{\prime}\varepsilon_{0}}, R_{\lambda^{\prime}\varepsilon_{1}}\}$
with linearly independent NBS wave functions 
$R_{\lambda^{\prime}\varepsilon_{0}}$ and 
$R_{\lambda^{\prime}\varepsilon_{1}}$.
%
For the spin
singlet
state, we extract the 
central potential as 
$
V_{\lambda \lambda^{\prime}}^{(Central)}(r;J=0)=
(\theta_{\lambda \lambda^{\prime}})^{-1}
({ R}^{-1})_{\varepsilon^{\prime}\lambda^{\prime}}
({\nabla^2\over 2\mu_{\lambda}}-{\partial\over \partial t})
{ R}_{\lambda\varepsilon^{\prime}}$. 
For the spin triplet state, 
the wave function 
is decomposed into 
the $S$- 
and 
$D$-wave components as 
\begin{equation}
 \left\{
 \begin{array}{l}
  R%
  (\vec{r};\ ^3S_1)={\cal P}R%
  (\vec{r};J=1)
   \equiv {1\over 24} \sum_{{\cal R}\in{ O}} {\cal R}
   R%
   (\vec{r};J=1),
   \\
  R%
  (\vec{r};\ ^3D_1)={\cal Q}R%
  (\vec{r};J=1)
   \equiv (1-{\cal P})R%
   (\vec{r};J=1).
 \end{array}
 \right.
\end{equation}
Therefore, 
the Schr\"{o}dinger equation with the LO 
potentials for the spin triplet state becomes
\begin{equation}
 \left\{\!\!\!
 \begin{array}{c}
  {\cal P} \\
  {\cal Q}
 \end{array}
 \!\!\!\right\}
 \!\times\!
 \left\{
  V^{(0)}_{\lambda \lambda^{\prime}}(r)
  \!+\!V^{(\sigma)}_{\lambda \lambda^{\prime}}(r)
  \!+\!V^{(T)}_{\lambda \lambda^{\prime}}(r)S_{12}
 \right\}
 \!\theta_{\lambda \lambda^{\prime}}
 { R}_{\lambda^{\prime}\varepsilon}(\vec{r},t-t_0)
 \!=
 \!\!\left\{\!\!\!
 \begin{array}{c}
  {\cal P} \\
  {\cal Q}
 \end{array}
 \!\!\!\right\}
 \!\times\!
 \left\{
     {\nabla^2\over 2\mu_{\lambda}} 
     -{\partial \over \partial t}
 \right\}
 \!{ R}_{\lambda\varepsilon}(\vec{r},t-t_0),
\end{equation}
from which
the 
central and tensor potentials, 
$V_{\lambda\lambda^{\prime}}^{(Central)}(r;J=0)=
(V^{(0)}(r)-3V^{(\sigma)}(r))_{\lambda\lambda^{\prime}}$ for $J=0$, 
$V_{\lambda\lambda^{\prime}}^{(Central)}(r;J=1)=
(V^{(0)}(r) +V^{(\sigma)}(r))_{\lambda\lambda^{\prime}}$, 
and $V_{\lambda\lambda^{\prime}}^{(Tensor)}(r)$ for $J=1$, can be
determined\footnote{
The potential is obtained from the NBS 
wave function at moderately large imaginary time; it would be 
$t-t_{0} \gg 1/m_{\pi} \sim 1.4$~fm. 
In addition, 
no single state saturation between the ground state 
and the excited states with respect to the relative motion, 
e.g., 
$t-t_{0} \gg (\Delta E)^{-1} = 
\left( (2\pi)^2/(2\mu (La)^2) \right)^{-1} \simeq 8.0$~fm, 
is required for the HAL QCD method\cite{HALQCD:2012aa}. 
}. 
%

\section{
Comprehensive lattice QCD calculation 
with almost physical quark masses}

$N_f=2+1$ gauge configurations at almost the physical quark masses are 
used; they are generated on $96^4$ lattice by employing the RG improved 
(Iwasaki) gauge action at $\beta=1.82$ with the nonperturbatively $O(a)$ 
improved Wilson quark (clover) action at 
$(\kappa_{ud},\kappa_{s})=(0.126117,0.124790)$ with $c_{sw}=1.11$ and 
the 6-APE stout smeared links with the smearing parameter $\rho=0.1$. 
Preliminary studies show that the physical volume is 
$(aL)^4\approx$(8.1fm)$^4$ with the lattice 
spacing $a\approx 0.085$fm and 
$(m_{\pi},m_{K})\approx(146,525)$MeV. 
See Ref.\cite{Ishikawa:2015rho} for details on the generation of 
the gauge configuration.
The periodic (Dirichlet) boundary condition is used for spacial (temporal) 
directions; wall quark source is employed with Coulomb gauge fixing which 
is separated from the Dirichlet boundary by $|t_{DBC}-t_{0}|=48$. 
Forward and backward propagation in time are combined by using the charge conjugation 
and time reversal symmetries to double the statistics. 
Each gauge configuration is used four times by using the hypercubic 
SO$(4,\mathbb{Z})$ symmetry of $96^4$ lattice. 
A large number of baryon-baryon 
potentials including the 
channels 
from $NN$ to $\Xi\Xi$ 
are studied 
by means of HAL QCD method\cite{DoiIshiiSasaki2017LAT}. 
See also Ref.\cite{Nemura:2015yha} for the thoroughgoing 
consistency check in the numerical outputs 
and comparison at various occasions 
between the UCA\cite{Doi:2012xd} and 
the present algorithm\cite{Nemura:2014eta}. 
In this report, 
96 wall sources 
are used for the 207 gauge configurations 
at every 10 trajectories. 
Statistical data are averaged with the bin size 23. 
Jackknife method is used to estimate the statistical errors.

%

\section{Results}

\subsection{Effective masses from single baryons' correlation function}
%
%
\begin{figure}[tb]
  \centering \leavevmode
  \sidecaption
  \includegraphics[width=0.4752\textwidth]{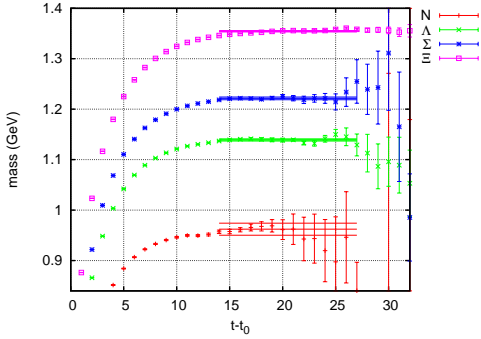}
  \caption{%
    The effective mass of single baryon's correlation functions 
    with utilising wall sources. %
    \label{Fig_Effmass}}
\end{figure}
%
%
In the following analysis to obtain the potential, 
we use 
the single baryon's correlation functions, 
$(C_{B_{1}}(t-t_{0})C_{B_{2}}(t-t_{0}))^{-1}$, 
instead of the 
simple exponential functional form ${\rm e}^{(m_{B_{1}}+m_{B_{2}})(t-t_{0})}$ 
in order to calculate the normalised four-point correlation function. 
It would be beneficial to reduce the statistical noise 
because of 
the statistical correlation between the numerator and the denominator 
in the normalised four-point correlation function.

Fig.~\ref{Fig_Effmass} shows the effective masses of 
the single baryon's correlation function. 
For the baryons $N,\Lambda$, and $\Sigma$, 
the plateaux start from the time slice around $t-t_{0} \approx 14$. 
Therefore it is favourable that the potentials are obtained at 
the time slices $t-t_{0} \gtrsim 14$. 
In this report we present 
preliminary results of potentials 
at 
time slices ($t-t_0=5-14$) of our on-going work.

\subsection{$\Sigma N$ ($I=3/2$) system}
\subsubsection{Potentials}
%
%
\begin{figure}[tb]
 \begin{minipage}[t]{0.33\textwidth}
  \centering \leavevmode
  \includegraphics[width=0.99\textwidth]{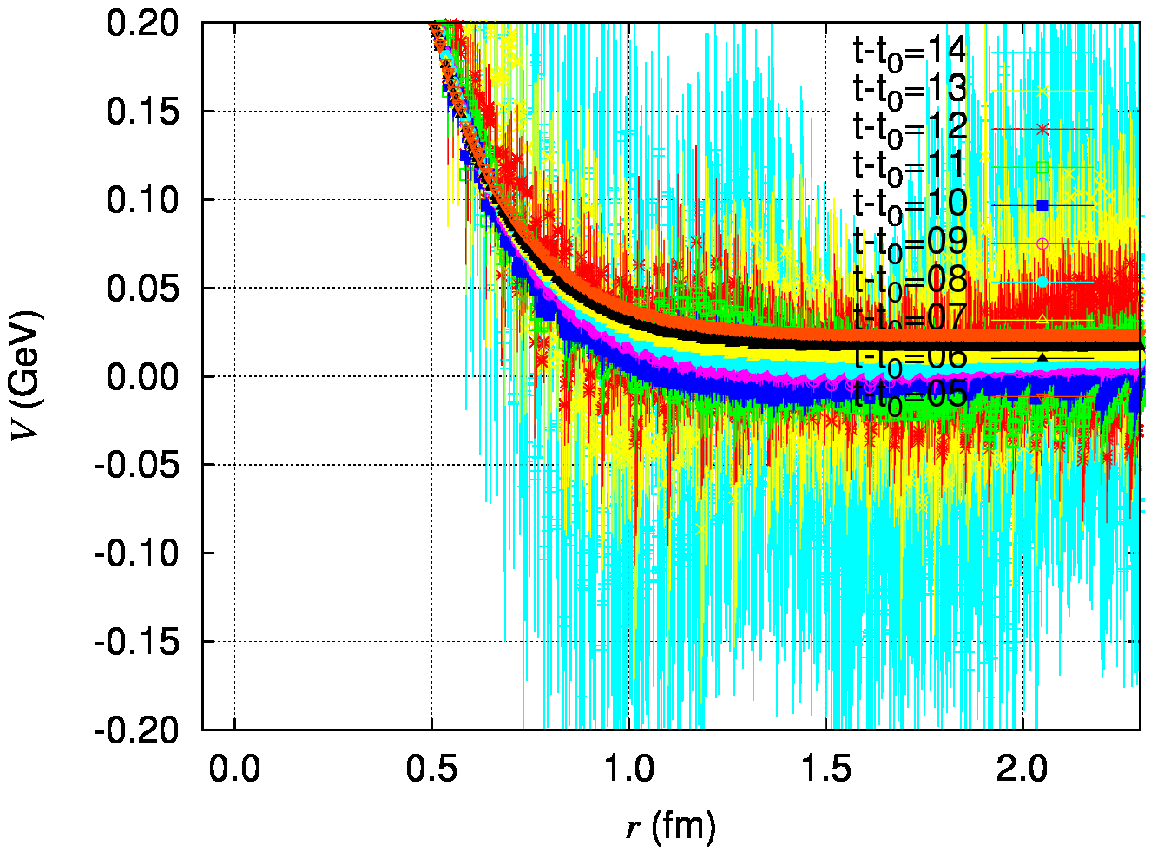}%
 \end{minipage}~
 \hfill
 \begin{minipage}[t]{0.33\textwidth}
  \centering \leavevmode
  \includegraphics[width=0.99\textwidth]{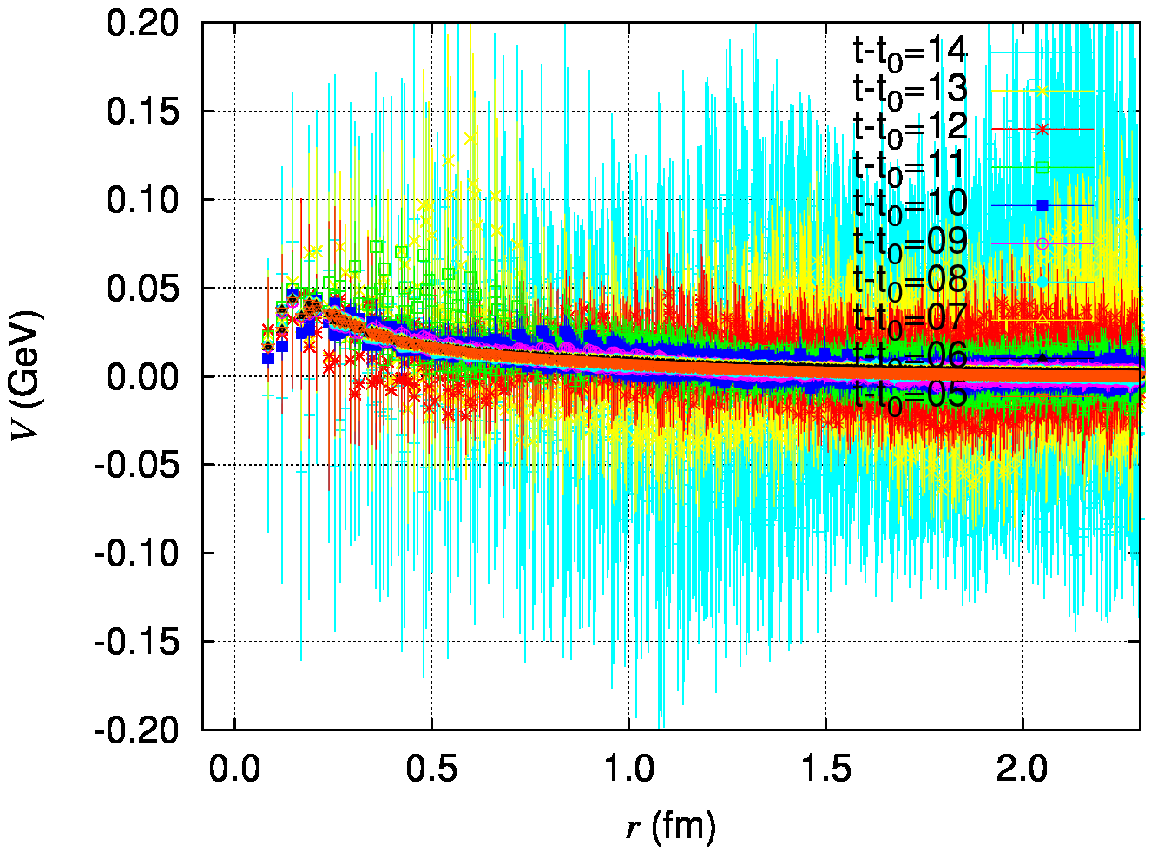}%
 \end{minipage}~
 \hfill
 \begin{minipage}[t]{0.33\textwidth}
  \centering \leavevmode
  \includegraphics[width=0.99\textwidth]{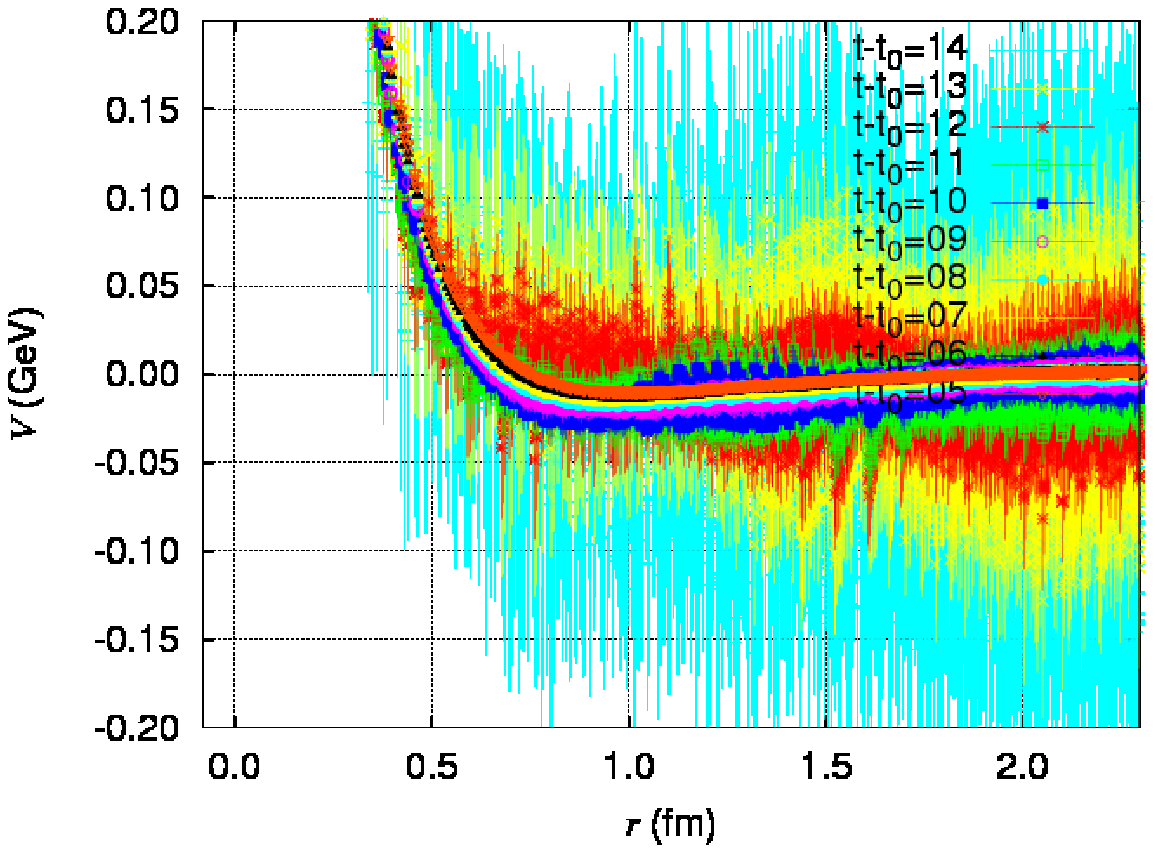}%
 \end{minipage}
 \caption{The $\Sigma N$ potentials of 
   $^3S_1-^3D_1$ central (left), 
   $^3S_1-^3D_1$ tensor  (centre), and 
   $^1S_0$ central (right) 
   in the $I=3/2$ channel. 
   \label{VC3E1_VT3E1_VC1S0_SN_2I3}}
\end{figure}
%
%
Fig.~\ref{VC3E1_VT3E1_VC1S0_SN_2I3} shows 
the central potential in the $^3S_1-^3D_1$ (left), 
the tensor  potential in the $^3S_1-^3D_1$ (centre), and 
the central potential in the $^1S_0$ (right) states 
of $\Sigma N$ ($I=3/2$) system, respectively. 
The stronger repulsive core of the central potential in the $^3S_1-^3D_1$ 
is seen in 
wider radial distance 
$r \lesssim 1$~fm; 
such a strong repulsion is 
consistent with quark model's prediction 
that is almost Pauli forbidden state 
in the flavor $\bm{10}$ representation. 
In addition, 
the central potential in the $^3S_1-^3D_1$ obtained at 
time slices $t-t_0 \ge 10$ shows small attractive well. 
The tensor potential is not as strong as the $NN$ tensor potential. 
The statistical fluctuation of the tensor potential becomes large at the time 
slices $t-t_0 \ge 11$ while 
that of the tensor potential 
at $t-t_0 \le 10$ 
does not. 
These observations are consistent with the scattering phase shift 
calculated below. 
On the other hand, for the $^1S_0$ state 
the repulsive core of the central potential is relatively short ranged; 
the attractive force is seen in medium to long distance. 
This behaviour is similar to the $NN$ $^1S_0$ 
because this state belongs to flavor $\bm{27}$.

\subsubsection{Scattering phase shifts}

%
\begin{figure}[htb]
 \begin{minipage}[t]{0.33\textwidth}
  \centering \leavevmode
  \includegraphics[width=0.99\textwidth]{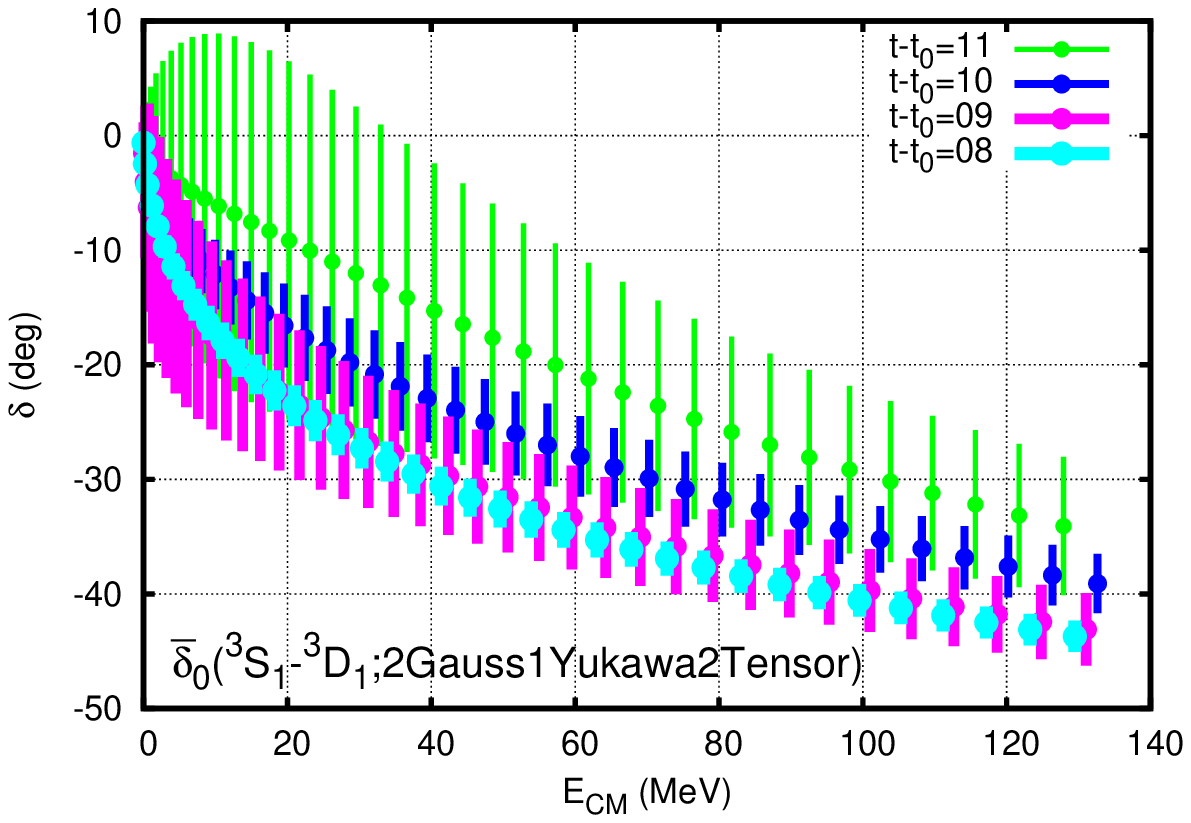}
 \end{minipage}~
 \hfill
 \begin{minipage}[t]{0.33\textwidth}
  \centering \leavevmode
  \includegraphics[width=0.99\textwidth]{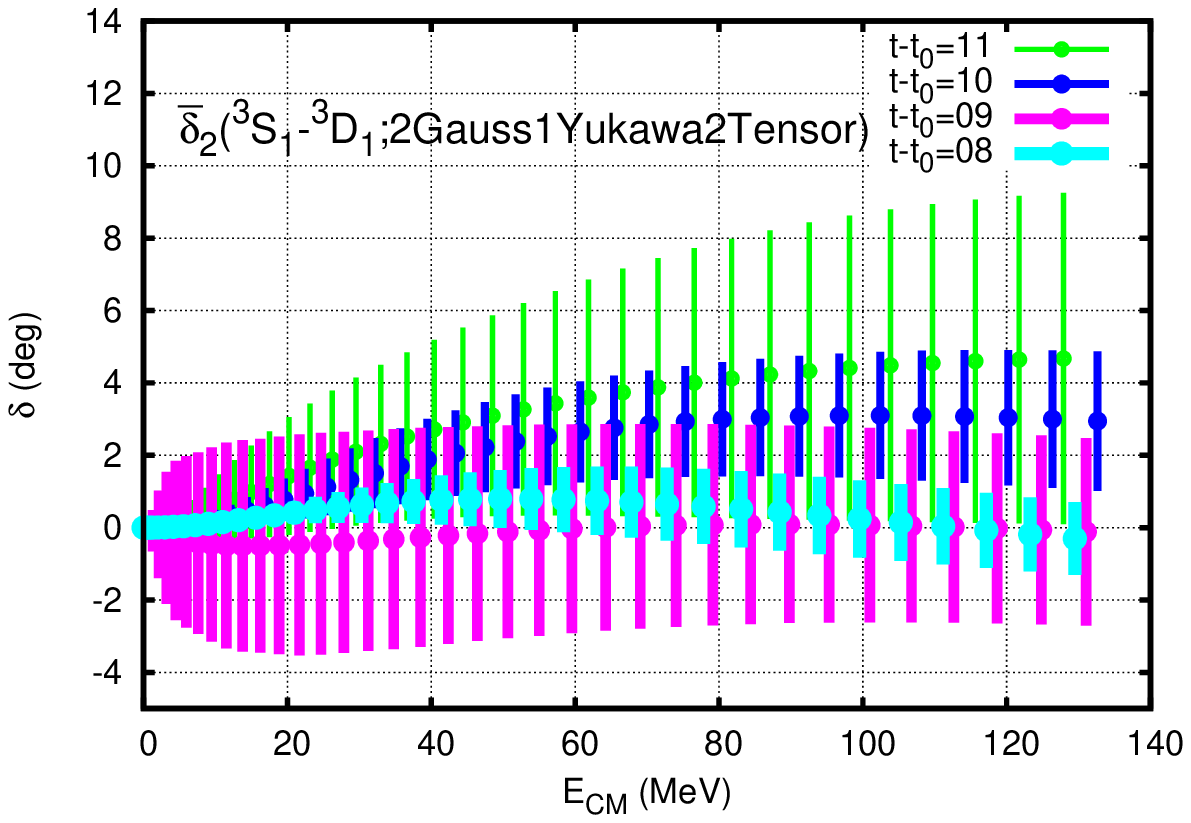}
 \end{minipage}~
 \hfill
 \begin{minipage}[t]{0.33\textwidth}
  \centering \leavevmode
  \includegraphics[width=0.99\textwidth]{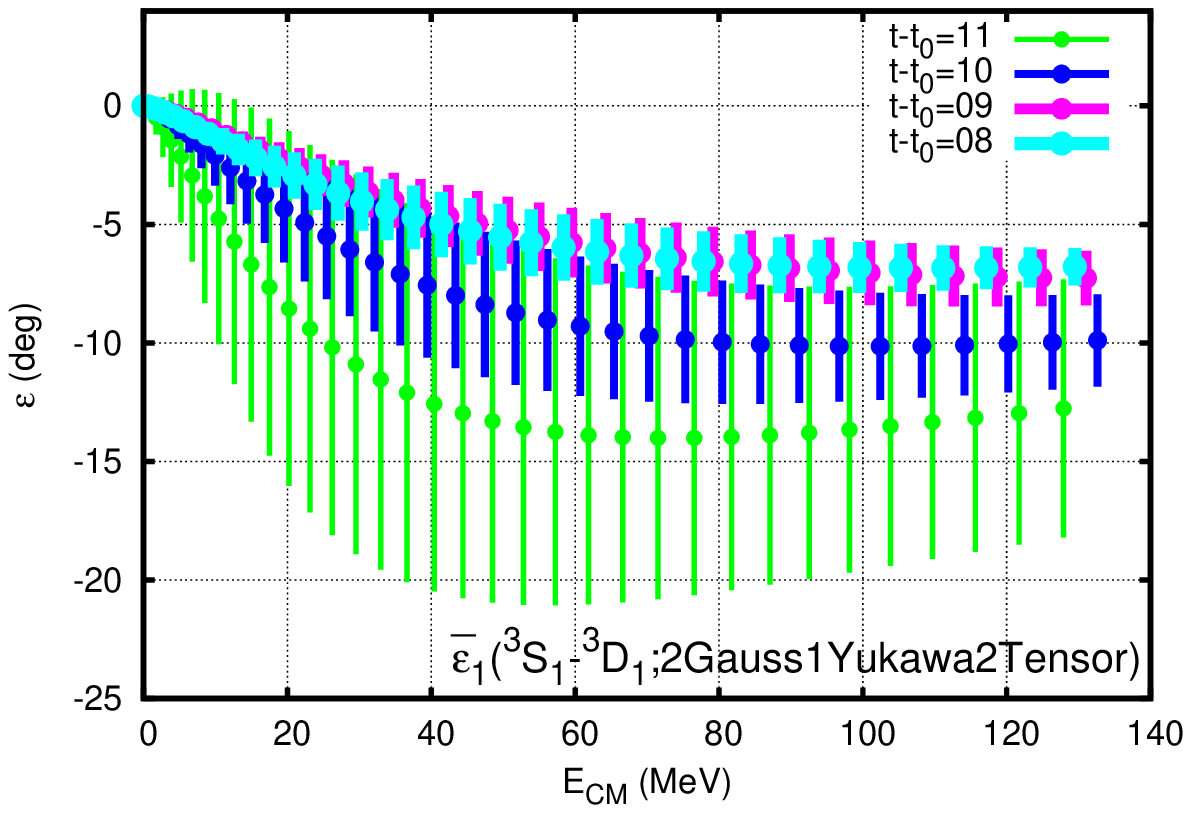}
 \end{minipage}
  \caption{Scattering bar-phase shifts and mixing angle in the 
    $^3S_1-^3D_1$ states of $I=3/2$ $\Sigma N$ system, 
    $\bar{\delta}_0$ (left), 
    $\bar{\delta}_2$ (centre), and 
    $\bar{\varepsilon}_1$ (right), 
    obtained from parametrised functional form Eq.~(\ref{VCndVT}) by 
    solving the Schr\"{o}dinger equation. 
    \label{Fig_Phsft3E1SN2I3}}
\end{figure}
%
%
The potential itself is not a physical observable. 
A reliable %
comparison 
with other results from experimental and/or theoretical (phenomenological) 
approaches 
should be made through physical observables, e.g., scattering phase shift. 
In order to obtain the scattering phase shift from 
present lattice QCD potential 
we first parametrise the potential %
with an analytic functional form. 
As the first attempt, we use following functional forms for the central and 
tensor potentials, %
respectively. 
\begin{equation}
  \begin{array}{l}
    V_{C}(r)
    = v_{C1} {\rm e}^{-\kappa_{C1} r^2}
    + v_{C2} {\rm e}^{-\kappa_{C2} r^2}
    + v_{C3}\left( 1-{\rm e}^{-\alpha_{C} r^2}\right)^2 
    \left( {{\rm e}^{-\beta_{C} r}\over r} \right)^2, 
    \\
    V_{T}(r)
    = v_{T1} \left( 1-{\rm e}^{-\alpha_{T1} r^2} \right)^{2} 
    \left( 1+{3\over \beta_{T1} r}+{3\over (\beta_{T1} r)^2} \right) 
         {{\rm e}^{-\beta_{T1} r}\over r}
    + v_{T2} \left( 1-{\rm e}^{-\alpha_{T2} r^2} \right)^{2} 
    \left( 1+{3\over \beta_{T2} r}+{3\over (\beta_{T2} r)^2} \right) 
              {{\rm e}^{-\beta_{T2} r}\over r}.
  \end{array}
  \label{VCndVT}
\end{equation}
Figure~\ref{Fig_Phsft3E1SN2I3} shows the scattering phase shifts in 
$^3S_1-^3D_1$ channels of $\Sigma N (I=3/2)$ system 
obtained by solving the Schr\"{o}dinger equation with 
above parametrised analytic functions. 
For the $^3S_1-^3D_1$ channels, 
the scattering matrix is parametrised with three real parameters 
bar-phase shifts and mixing angle~\cite{Stapp:1956mz}:
\begin{equation}
S = 
  \left( \begin{array}{cc}
    {\rm e}^{i \bar{\delta}_{J-1}} & 0 \\ %
    0 & {\rm e}^{i \bar{\delta}_{J+1}}    %
  \end{array} \right)
  \left( \begin{array}{cc}
      \cos 2 \bar{\varepsilon}_{J} & i \sin 2 \bar{\varepsilon}_{J} \\
    i \sin 2 \bar{\varepsilon}_{J} &   \cos 2 \bar{\varepsilon}_{J} 
  \end{array} \right)
  \left( \begin{array}{cc}
    {\rm e}^{i \bar{\delta}_{J-1}} & 0 \\ %
    0 & {\rm e}^{i \bar{\delta}_{J+1}}    %
  \end{array} \right).
\end{equation}
The phase shift $\bar{\delta}_0$ at the time slices $t-t_0=9-11$ shows 
the interaction is repulsive 
while the phase shift $\bar{\delta}_2$ behaves around almost zero degree. 
%
%
\begin{figure}[hbt]
  \centering \leavevmode
  \sidecaption
  \includegraphics[width=0.3267\textwidth]{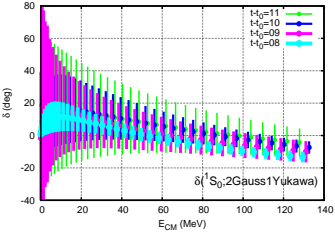}
 \footnotesize
 \caption{Scattering phase shift in the $^1S_0$ state 
   of $I=3/2$ $\Sigma N$ system, 
   obtained from parametrised functional form Eq.~(\ref{VCndVT}) by 
   solving the Schr\"{o}dinger equation. 
   \label{Fig_Phsft1S0SN2I3}}
\end{figure}
%
%
Figure~\ref{Fig_Phsft1S0SN2I3} shows the scattering phase shift in $^1S_0$ 
channel of $\Sigma N (I=3/2)$ system obtained through 
the above parametrised functions. 
The present result shows that the interaction in the $^1S_0$ channel 
of $\Sigma N (I=3/2)$ system is attractive on average 
though the fluctuation is 
large especially for the time slices $t-t_0=9,11$. 
The lattice potentials at flavor 
$SU(3)$ limit~\cite{Inoue:2011ai} show that 
group theoretical classification based on quark model 
works for clarifying the general behaviour of 
various baryon-baryon interactions in the $S$-wave; 
the $\Sigma N$ $I=3/2$ $^3S_1-^3D_1$ belongs to $\bm{10}$ which 
is almost Pauli forbidden while 
the $\Sigma N$ $I=3/2$ $^1S_0$ belongs to $\bm{27}$ which is same as 
$NN$ $^1S_0$. 
The present $S$-wave (dominated) phase shifts, 
the repulsive (attractive) behaviour of $\bar{\delta}_0$ ($\delta(^1S_0)$), 
augur well for future quantitative conclusions with larger statistics. 
Incidentally, these behaviours are also qualitatively similar to 
recent 
studies~\cite{Beane:2012ey,Fujiwara:1996qj,Arisaka:2000vu,Haidenbauer:2013oca}. 
For both Figs.~\ref{Fig_Phsft3E1SN2I3} and \ref{Fig_Phsft1S0SN2I3}, 
the parametrisation procedure through the functional form 
may not be so stable at this moment especially for $t-t_0=11$. 
The present phase shifts and mixing angle should be regarded as 
preliminary results so that the large errorbars would be 
improved by future analysis with larger statistical data. 

%

%

%

\subsection{$\Lambda N-\Sigma N$ ($I=1/2$) coupled-channel systems}
%
%
\begin{figure}[tb]
 \begin{minipage}[t]{0.33\textwidth}
  \centering \leavevmode
  \includegraphics[width=0.99\textwidth]{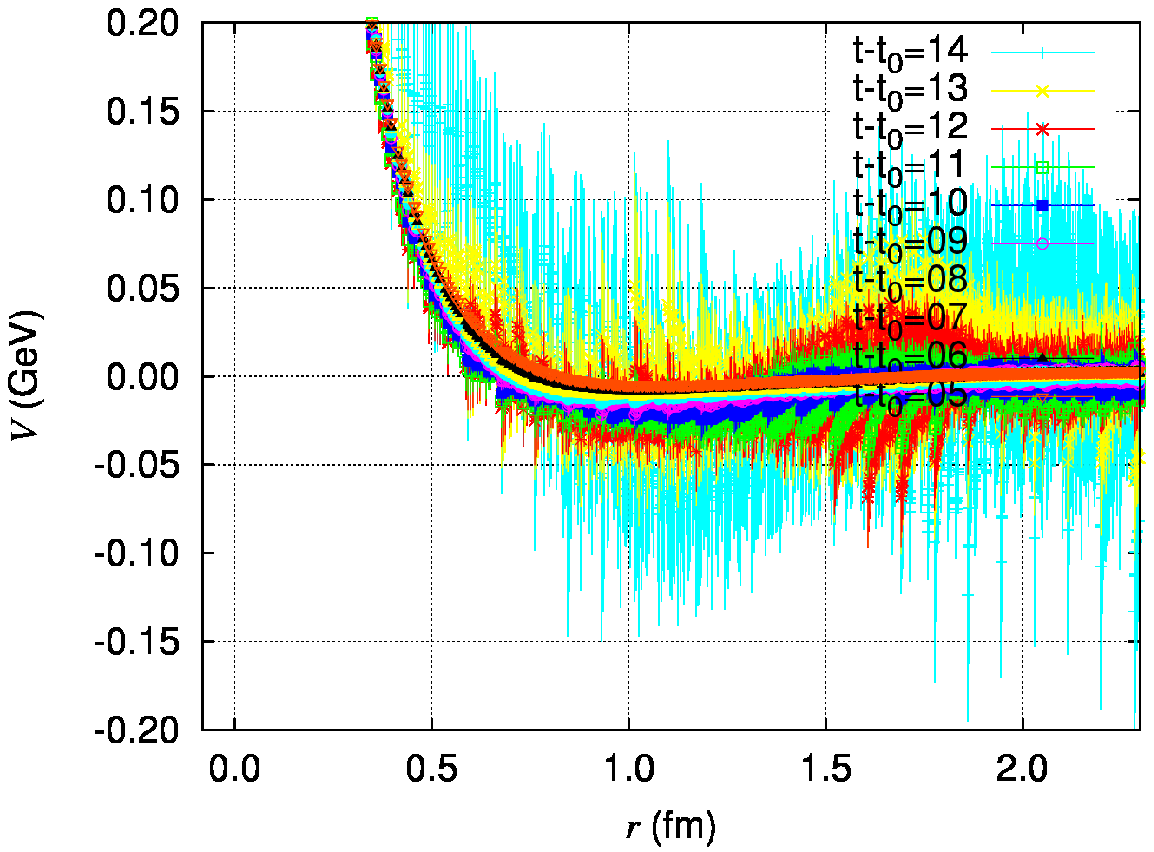}%
 \end{minipage}~
 \hfill
 \begin{minipage}[t]{0.33\textwidth}
  \centering \leavevmode
  \includegraphics[width=0.99\textwidth]{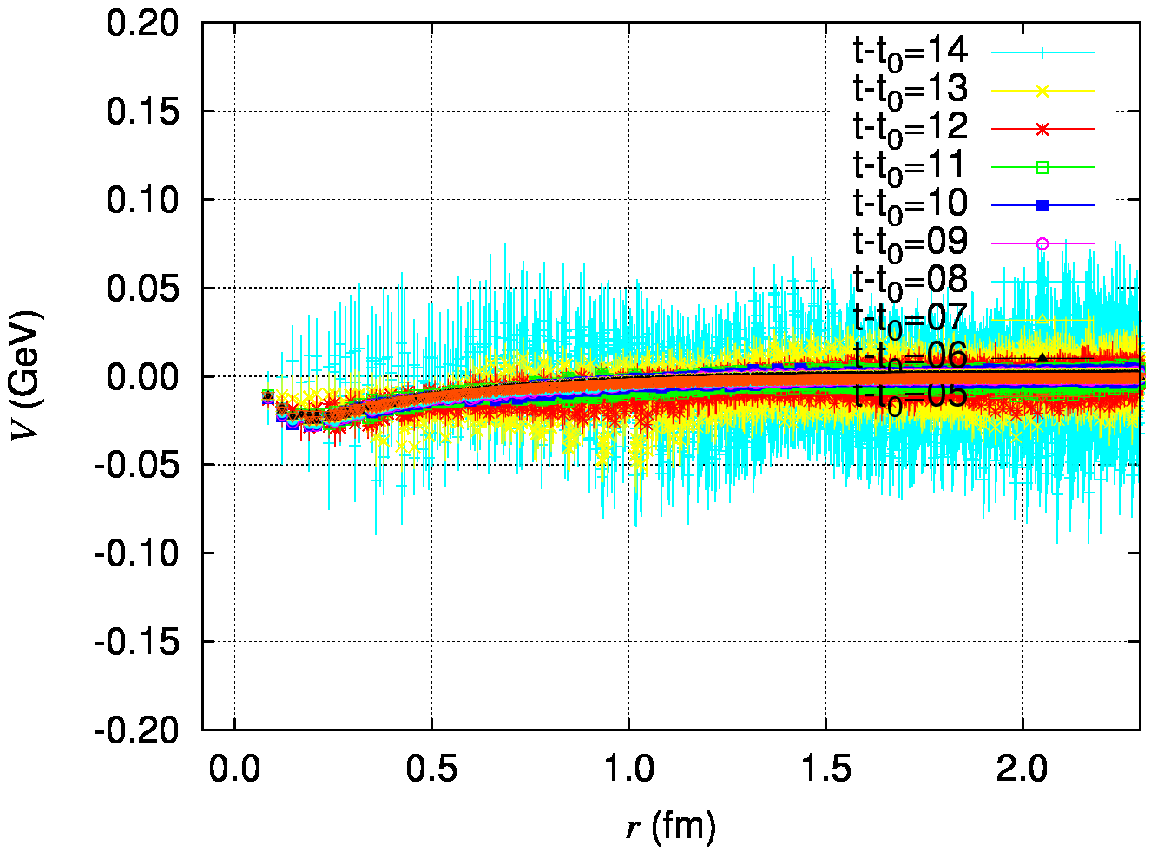}%
 \end{minipage}~
 \hfill
 \begin{minipage}[t]{0.33\textwidth}
  \centering \leavevmode
  \includegraphics[width=0.99\textwidth]{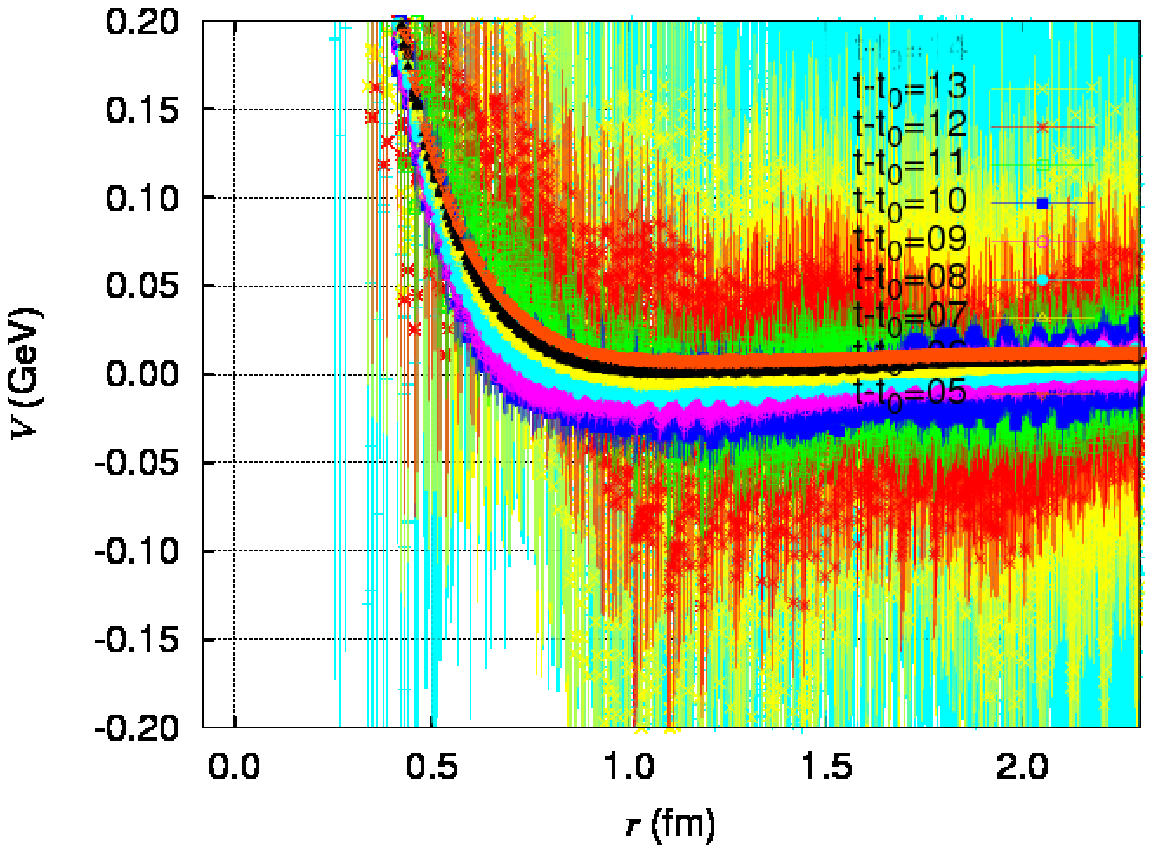}%
 \end{minipage}
 \caption{The $\Lambda N-\Lambda N$ potentials for 
   $^3S_1-^3D_1$ central (left), 
   $^3S_1-^3D_1$ tensor  (centre), and 
   $^1S_0$ central (right). 
   \label{VC3E1_VT3E1_VC1S0_LN}}
\end{figure}
%
%
Fig.~\ref{VC3E1_VT3E1_VC1S0_LN} shows $\Lambda N-\Lambda N$ diagonal part 
for 
the central potential in the $^3S_1-^3D_1$ (left),
the tensor  potential in the $^3S_1-^3D_1$ (centre), and 
the central potential in the $^1S_0$ (right) states 
of $\Lambda N-\Sigma N$ ($I=1/2$) system, respectively. 
There are repulsive cores in the short distance region and medium to long range 
attractive well for both central potentials. 
The relatively weak tensor potential is found. 
%
%
\begin{figure}[tb]
 \begin{minipage}[t]{0.33\textwidth}
  \centering \leavevmode
  \includegraphics[width=0.99\textwidth]{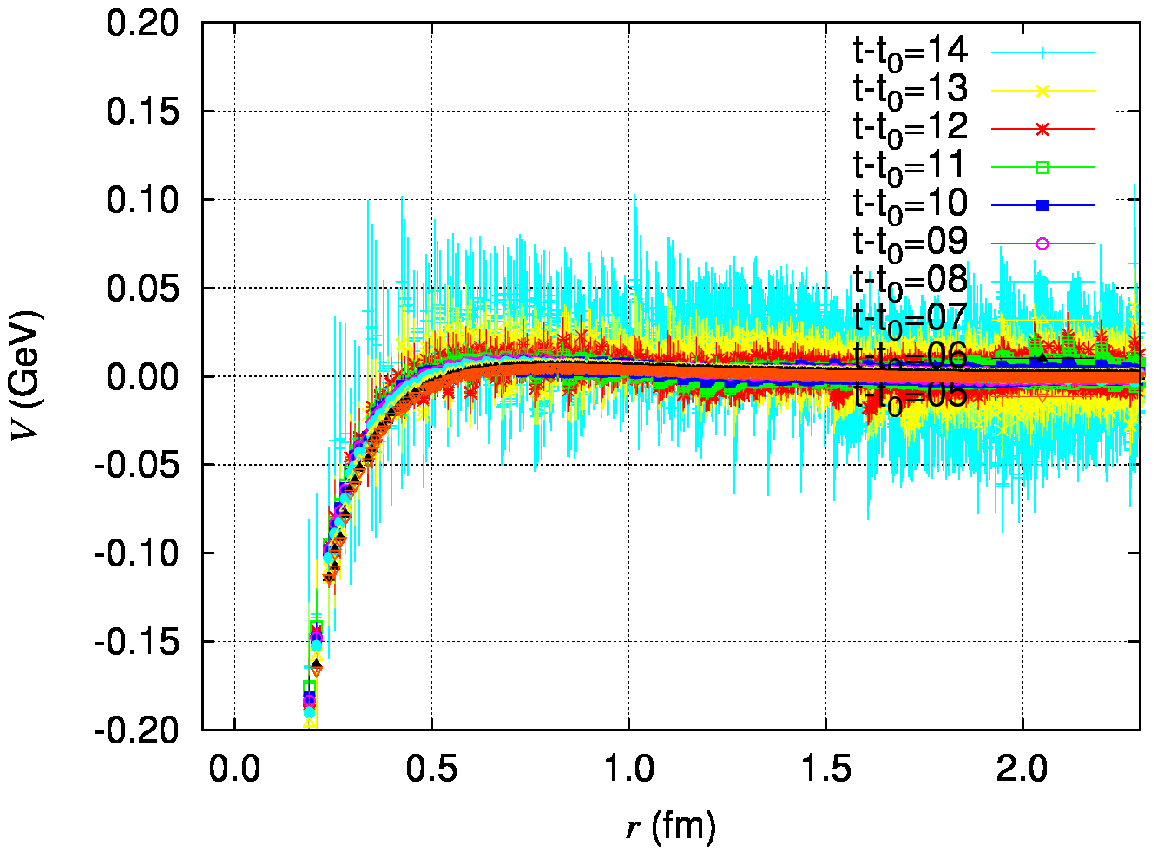}%
 \end{minipage}~
 \hfill
 \begin{minipage}[t]{0.33\textwidth}
  \centering \leavevmode
  \includegraphics[width=0.99\textwidth]{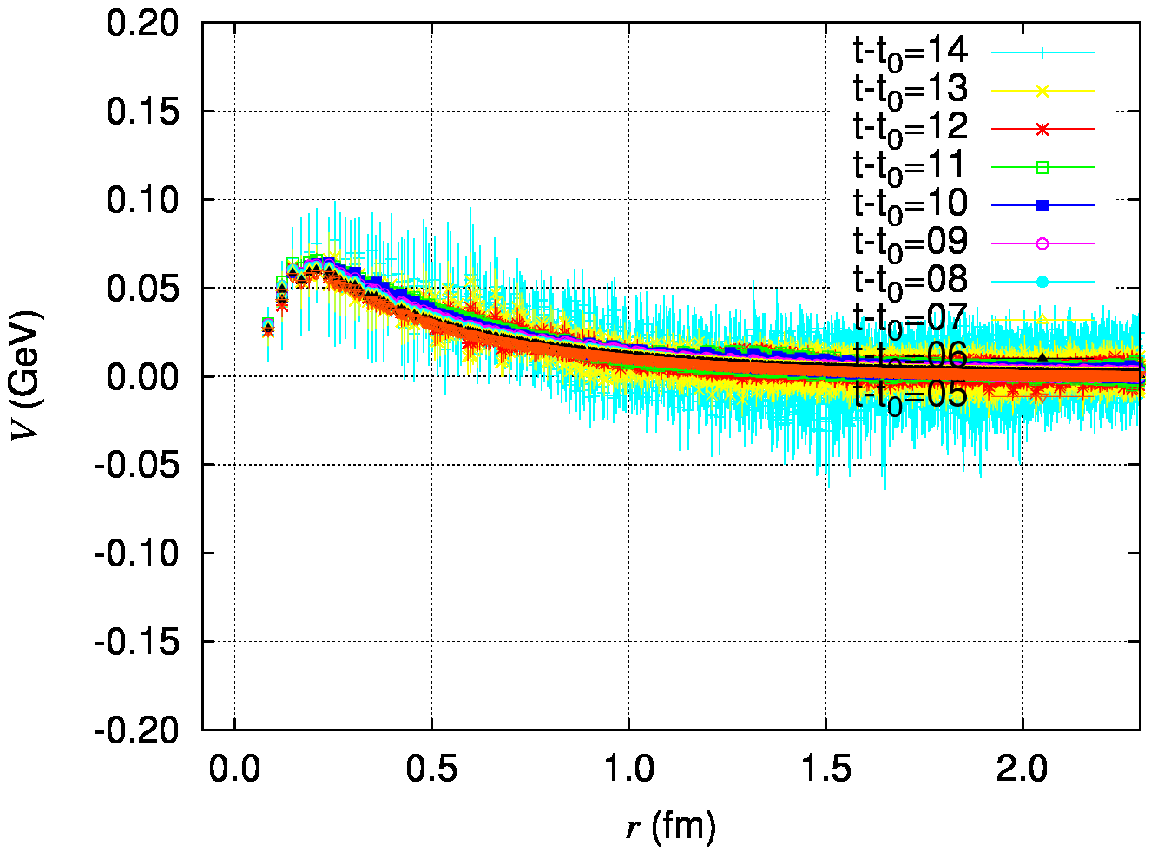}%
 \end{minipage}~
 \hfill
 \begin{minipage}[t]{0.33\textwidth}
  \centering \leavevmode
  \includegraphics[width=0.99\textwidth]{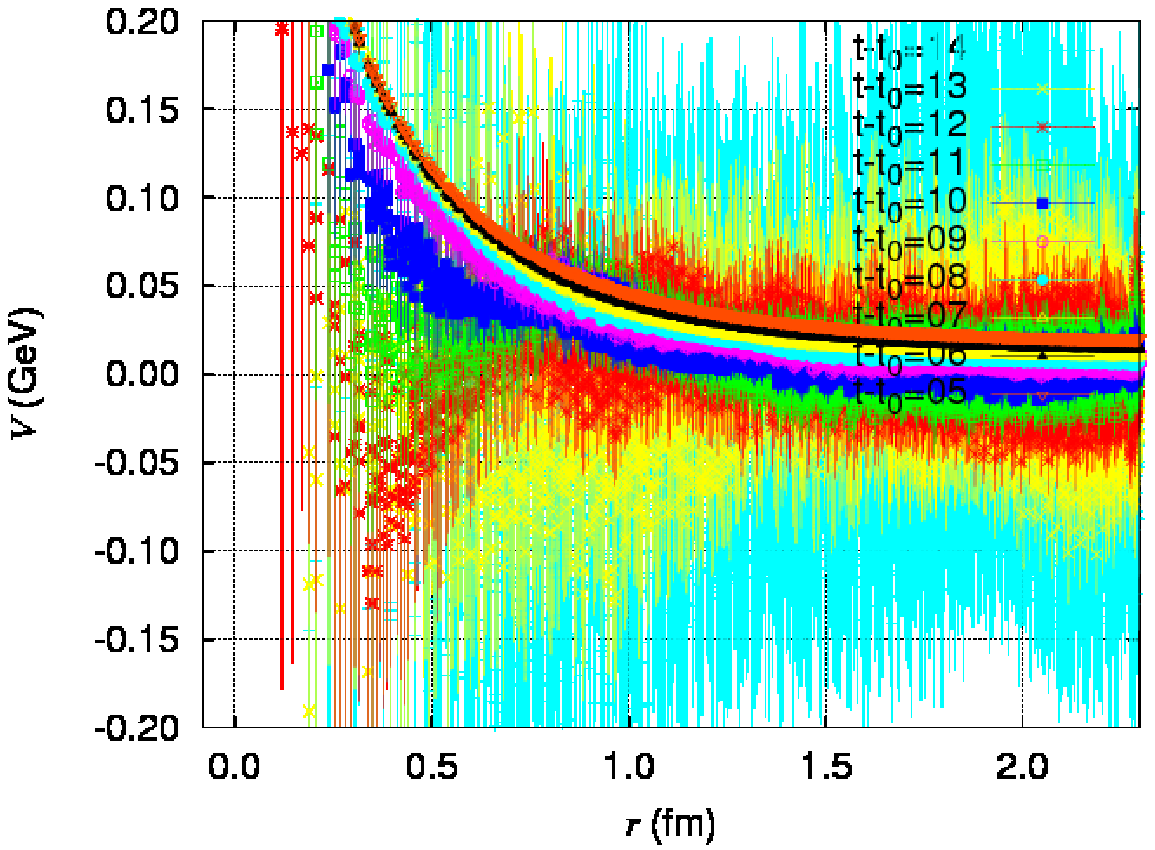}%
 \end{minipage}
 \caption{The $\Lambda N\rightarrow\Sigma N$ potentials for 
   $^3S_1-^3D_1$ central (left), 
   $^3S_1-^3D_1$ tensor  (centre), and 
   $^1S_0$ central (right). 
   \label{VC3E1_VT3E1_VC1S0_LNtoSN}}
\end{figure}
%
%
Fig.~\ref{VC3E1_VT3E1_VC1S0_LNtoSN} shows $\Lambda N\rightarrow\Sigma N$ 
transition part for 
the central potential in the $^3S_1-^3D_1$ (left),
the tensor  potential in the $^3S_1-^3D_1$ (centre), and 
the central potential in the $^1S_0$ (right) states 
of $\Lambda N-\Sigma N$ ($I=1/2$) system, respectively. 
The $^3S_1-^3D_1$ central potential is found to be short ranged. 
The tensor potential is not as strong as the $NN$ tensor potential but 
it has sizable strength. 
The statistical fluctuation in the $^1S_0$ central potential is still large. 
%
%
\begin{figure}[tb]
 \begin{minipage}[t]{0.33\textwidth}
  \centering \leavevmode
  \includegraphics[width=0.99\textwidth]{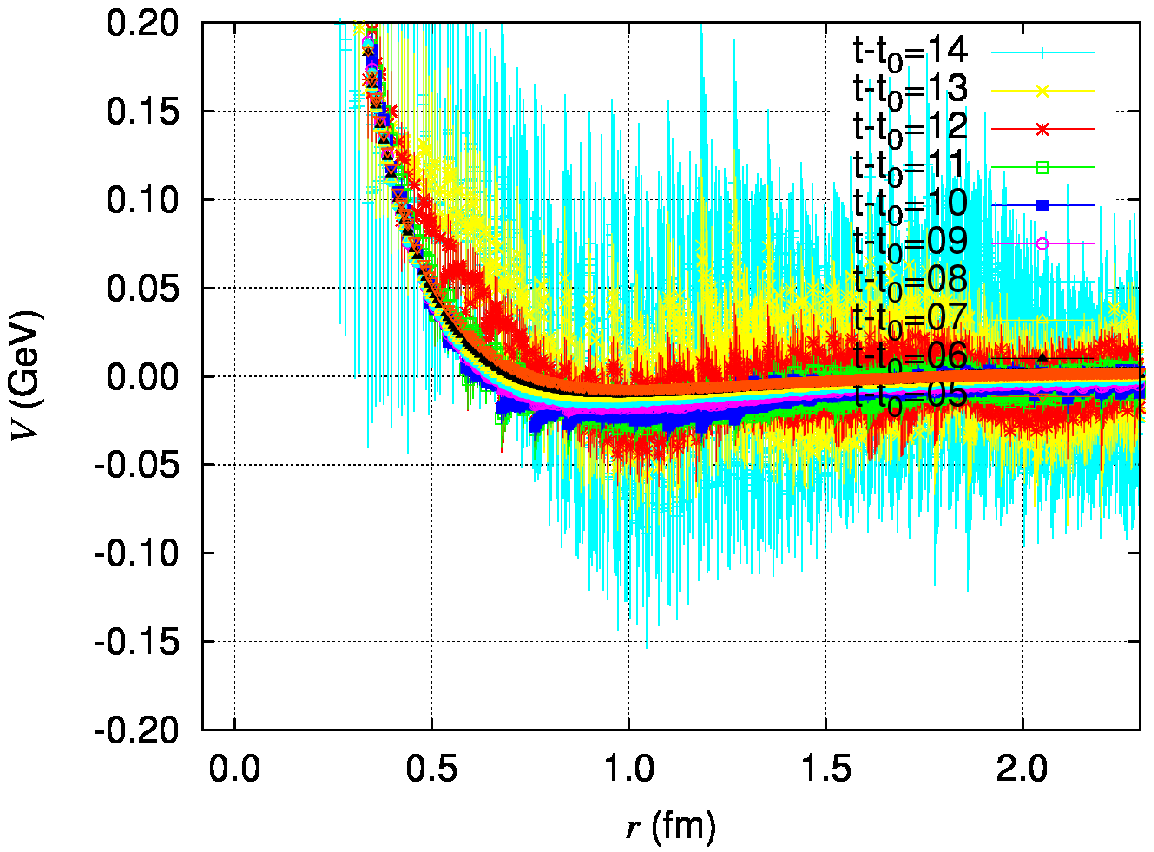}%
 \end{minipage}~
 \hfill
 \begin{minipage}[t]{0.33\textwidth}
  \centering \leavevmode
  \includegraphics[width=0.99\textwidth]{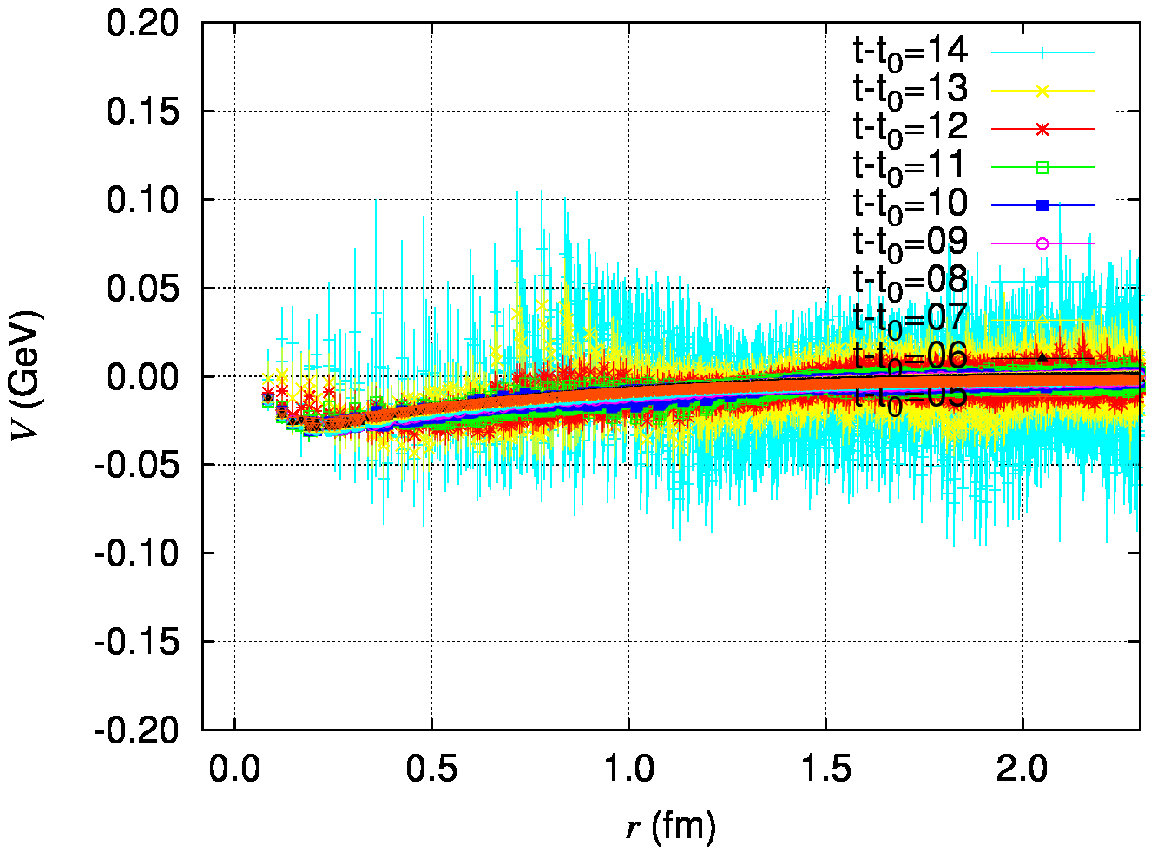}%
 \end{minipage}~
 \hfill
 \begin{minipage}[t]{0.33\textwidth}
  \centering \leavevmode
  \includegraphics[width=0.99\textwidth]{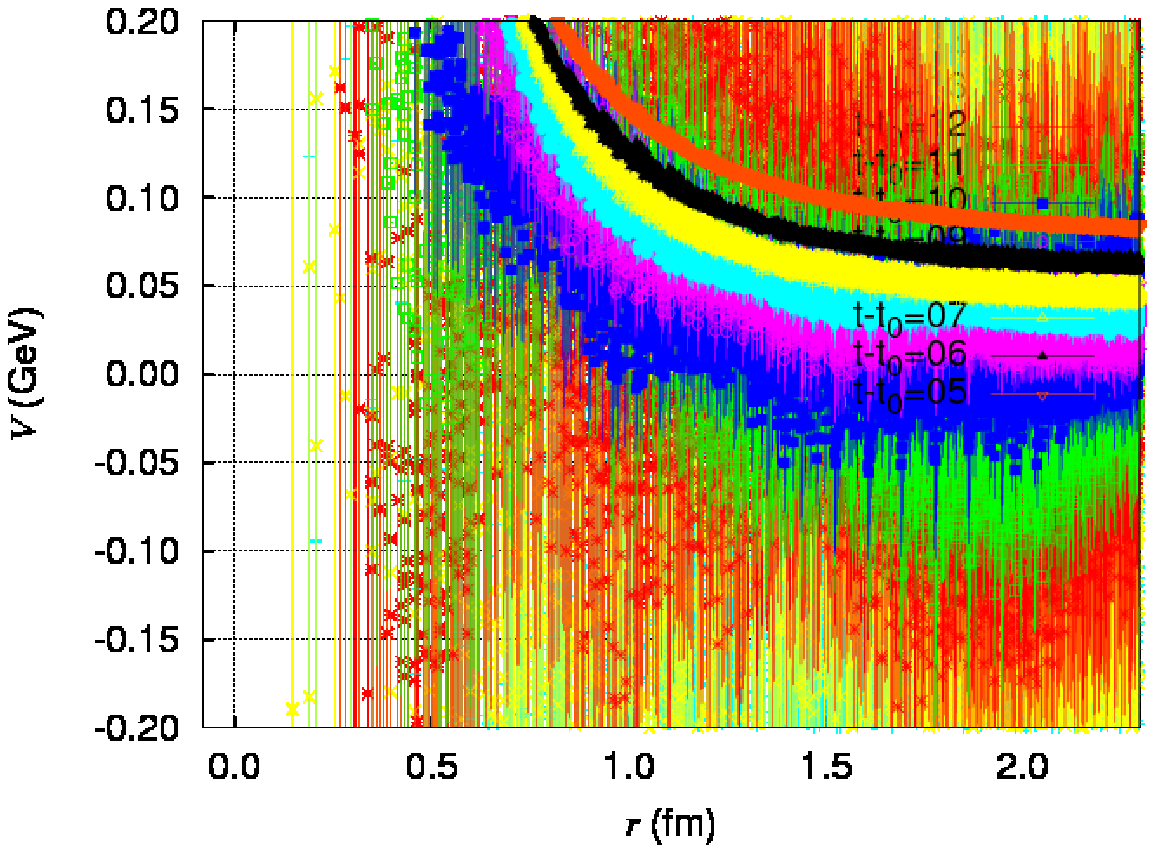}%
 \end{minipage}
 \caption{The $\Sigma N-\Sigma N$ potentials of 
   $^3S_1-^3D_1$ central (left), 
   $^3S_1-^3D_1$ tensor  (centre), and 
   $^1S_0$ central (right) 
   in the $I=1/2$ channel. 
   \label{VC3E1_VT3E1_VC1S0_SN_2I1}}
\end{figure}
%
%
Fig.~\ref{VC3E1_VT3E1_VC1S0_SN_2I1} shows $\Sigma N-\Sigma N$ diagonal 
part for 
the central potential in the $^3S_1-^3D_1$ (left), 
the tensor  potential in the $^3S_1-^3D_1$ (centre), and 
the central potential in the $^1S_0$ (right) states 
of $\Lambda N-\Sigma N$ ($I=1/2$) system, respectively. 
There are short range repulsive core and medium range attractive well in 
the $^3S_1-^3D_1$ central potential. 
The very strong repulsive core is seen in the $^1S_0$ central potential; 
it could be due to the large contribution of flavor $\bm{8}_s$ component, 
where we have 
$|\Sigma N\rangle ={1\over\sqrt{10}}(3|\bm{8}_{s}\rangle-|\bm{27}\rangle)$ 
in the flavor SU(3) limit. 
The statistical fluctuation 
in the repulsive channel seems to be large. 
\section{Summary}

In this report, the preliminary results of the $\Lambda N$, $\Sigma N$ and 
their coupled-channel potentials are presented. 
For the $\Sigma N$ ($I=3/2$) interaction, 
phase shifts are calculated for the $^3S_1-^3D_1$ and $^1S_0$ states. 
The phase shift $\bar{\delta}_{0}$ in the $^3S_1-^3D_1$ channel 
shows that the $\Sigma N$ ($I=3/2$,$^3S_1$) interaction is repulsive. 
The phase shift in the $\Sigma N$ ($I=3/2$,$^1S_0$) channel shows 
that the interaction is attractive on average. 
These results are qualitatively consistent with recent 
phenomenological approaches. 
For the $\Lambda N-\Sigma N$ coupled-channel system, 
the potentials in the $^1S_0$ channel have still large 
statistical fluctuations because the number of statistics 
in the spin-singlet is factor 3 smaller than the number of statistics 
in the spin-triplet. 
In addition, large contribution from flavor $\bm{8}_{s}$ component in the 
$\Sigma N$ ($I=1/2$, $^1S_0$) could deteriorate the signal in the 
$\Sigma N$ $^1S_0$ potential. 
Further calculations to obtain physical quantities 
with increased statistics are in progress and 
will be reported elsewhere.

\bigskip

\begin{acknowledgement}
We thank all collaborators in this project, above all, 
members of PACS Collaboration for the gauge configuration generation. 
The lattice QCD calculations have been performed 
on the K computer at RIKEN, AICS 
(hp120281, hp130023, hp140209, hp150223, hp150262, hp160211, hp170230),
HOKUSAI FX100 computer at RIKEN, Wako (
G15023, G16030, G17002)
and HA-PACS at University of Tsukuba 
(14a-25, 15a-33, 14a-20, 15a-30).
We thank ILDG/JLDG~\cite{ILDGJLDG}
which serves as an essential infrastructure in this study.
This work is supported in part by 
MEXT Grant-in-Aid for Scientific Research 
(JP16K05340, JP25105505),
and SPIRE (Strategic Program for Innovative Research) Field 5 project and 
``Priority issue on Post-K computer'' (Elucidation of the Fundamental Laws
and Evolution of the Universe) and 
Joint Institute for Computational Fundamental Science (JICFuS).
\end{acknowledgement}


%

\end{document}